\documentclass[12pt]{article}
\usepackage{graphicx}
\usepackage{amsmath}

\newtheorem{theorem}{Theorem}

\newtheorem{definition}[theorem]{Definition}

\input{tcilatex}

\begin{document}

\title{Special Relativity: Einstein's Spherical Waves versus Poincar\'{e}'s
Ellipsoidal Waves}
\author{Dr. Yves Pierseaux \\
%EndAName
{\small Physique des particules, Universit\'{e} Libre de Bruxelles (ULB) }\\
{\small ypiersea@ulb.ac.be}\\
{\small Talk given September 5, 2004, PIRT (London, Imperial College)}}
\maketitle

\begin{abstract}
We show that the image by the Lorentz transformation of a spherical
(circular) light wave, emitted by a moving source, is not a\ spherical
(circular) wave but an ellipsoidal (elliptical) light wave. Poincar\'{e}'s
ellipsoid (ellipse) is the direct geometrical representation of
Poincar\'{e}'s relativity of simultaneity. Einstein's spheres (circles) are
the direct geometrical representation of Einstein's convention of
synchronisation. Poincar\'{e} adopts another convention for the definition
of space-time units involving that the Lorentz transformation of an unit of
length is \textbf{directly proportional} to Lorentz transformation of an
unit of time. Poincar\'{e}'s relativistic kinematics predicts both a
dilation of time and an \textbf{expansion of space} as well.
\end{abstract}

\section{\protect\bigskip Introduction: Einstein's Spherical Wavefront%
\newline
\& Poincar\'{e}'s Ellipsoidal Wavefront}

Einstein writes in 1905, in the third paragraph of his famous paper :

\begin{quotation}
At the time $t=\tau =0$, when the origin of the two coordinates (K and k) is
common to the two systems, let a \textbf{spherical wave }be emitted
therefrom, and be propagated -with the velocity c in system K. If x, y, z be
a point just attained by this wave, then

\begin{equation}
x^{2}+y^{2}+z^{2}=c^{2}t^{2}
\end{equation}

Transforming this equation with our equations of transformation (see
Einstein's LT, 29), we obtain after a simple calculation

\begin{equation}
\xi ^{2}+\eta ^{2}+\zeta ^{2}=c^{2}\tau ^{2}
\end{equation}

The wave under consideration is therefore no less a \textbf{spherical wave }%
with velocity of propagation c when viewed in the moving system k. \cite{2}
\end{quotation}

\bigskip Poincar\'{e} writes in 1908 in his second paper on ''La dynamique
de l'\'{e}lectron'' with the subtitle ''Le principe de relativit\'{e}'':

\begin{quotation}
Imagine an observer and a source involved together in the transposition. The
wave surfaces emanating for the source will be \textbf{spheres}, having as
centre the successive positions of the source. The distance of this centre
from the present position of the source will be proportional to the time
elapsed since the emission - that is to say, to the radius of the sphere.
But for our observer, on account of the contraction, all these spheres will
appear as \textbf{elongated ellipsoids}. The compensation is now exact, and
this is explained by Michelson's experiments. \cite{14}
\end{quotation}

We can further find in Poincar\'{e}'s text the equation (two dimensions) of
an elongated light ellipse whose observer at rest (let us call him: O) is
situated at the centre and whose\ source S (with ''our observer'', let us
call him O') in moving is situated at the focus F of the ellipse.

The contrast between both great relativists, Einstein et Poincar\'{e}, about
an experiment that seems to be the same (\textit{the image of a spherical
wave emitted by a moving source}) is very clear: according to Einstein, the
image of a spherical wave is a spherical wave whilst according to
Poincar\'{e} the image of a spherical wave (around O) is an ellipsoidal
wave. Does the latter not know the invariance of the quadratic form? Not at
all because he does demonstrate, with the structure of group, in his first
paper on ,''La dynamique de l'\'{e}lectron'' \cite{11}, that the Lorentz
Transformation (\textbf{LT}) ''doesn't modify the quadratic form\textit{\ }$%
x^{2}+y^{2}+z^{2}-c^{2}t^{2}".$ \ We must point out \ that Poincar\'{e}'s
lengthened light waves has been completely ignored for a whole century by
the scientific community \footnote{%
Poincar\'{e}'s ellipsoidal wavefront was in fact mentioned in his course of
1905-1906 ''Les limites de la loi de Newton'' \cite{12}. We also find them
in ''La M\'{e}canique Nouvelle'' (1909) \cite{15}. In fact it was in 1904 at
a talk in Saint Louis that Poincar\'{e} first introduced the elongated
ellipsoidal wavefront as an \textit{alternative} and not as a \textit{%
consequence} of the contraction of the unit of length \cite{10}.}. We note
also that Poincar\'{e} doesn't use LT in the previous quotation and directly
deduces the ellipsoidal shape of the light wavefront from the principle of
contraction of (the unit) of length (see conclusion).

So who is right: Einstein or Poincar\'{e}? The best thing that we can do, to
solve this dilemma, is to apply a LT to a spherical wavefront.

\section{Image by LT of the Object ''Circular Wave''}

What is the image (the shape) in K of a spherical wave emitted in $t^{\prime
}=t=0$ by a source S at rest in the origin O' of K'? The LT defined by
Poincar\'{e} is:

\begin{equation}
x^{\prime }=k(x-\varepsilon t)\qquad \qquad y^{\prime }=y\qquad \qquad
t^{\prime }=k(t-\varepsilon x)
\end{equation}

We keep Poincar\'{e}'s notations where $\varepsilon ,k$ correspond to
Einstein-Planck's notations $\beta ,\gamma $ because, according to
Poincar\'{e} in his 1905 work about the theory of relativity,\textit{\ ''I
shall choose the units of length and of time in such a way that the velocity
of light is equal to unity'' }\cite{11}. The deep meaning of Poincar\'{e}'s
choice of space-time units\textit{\ with }$c=1$ will be specified in the
conclusion. In order to have \textit{one only} wavefront, we have to define
a time $t^{\prime }$ as unit of time $1_{t^{\prime }}$. The equation of the
circular wave front in K' (the geometrical locus of the object-points in K')
in $t^{\prime }=1$ is :

\begin{equation}
x^{\prime 2}+y^{\prime 2}=t^{\prime 2}=1_{t^{\prime }}
\end{equation}

%\FRAME{fhFU}{328.75pt}{300.0625pt}{0pt}{\Qcb{Image-Points (A, B, C...) in K
%of the Object-Points (A', B', C'...) of a spherical wavefront emitted by a
%source in K' ($\protect\varepsilon \sim 0.9,$ $k\sim 2.3$ $and$ $k\protect%
%\varepsilon \sim 2)$}}{}{Figure }{\special{language "Scientific Word";type
%"GRAPHIC";display "USEDEF";valid_file "T";width 328.75pt;height
%300.0625pt;depth 0pt;original-width 759.75pt;original-height
%667.0625pt;cropleft "0";croptop "1";cropright "1";cropbottom
%"0";tempfilename 'I6795Y00.wmf';tempfile-properties "XPR";}}

The unprimed coordinates of the image-points are given by the inverse LT:

\begin{equation}
x=k(x^{\prime }+\varepsilon t^{\prime })\qquad \qquad \qquad y=y^{\prime
}\qquad \qquad t=k(t^{\prime }+\varepsilon x^{\prime })
\end{equation}
The coordinates $(0,0,1)$ in K of the source in $t^{\prime }=1$ are $%
(k\varepsilon ,0,k)$ and $(k\varepsilon t^{\prime },0,kt^{\prime })$ in $%
t^{\prime }\neq 1$.

Let us determine the images (x, y, t) in K of different object-points in $%
t^{\prime }=1:$ $(1,0,1),$ $(-1,0,1),(0,1,1),(\frac{\sqrt{2}}{2},\frac{\sqrt{%
2}}{2},1)$ $etc.$ (see \textbf{figure 1 }in annex). The image-point E, $%
k(1+\varepsilon ),0,k(1+\varepsilon ),$ is on the large dotted circle $%
x^{2}+y^{2}=t^{2}$ with radius $r=t=k(1+\varepsilon ).$

The image-point A, $k(\varepsilon -1),0,k(1-\varepsilon ),$ is on the small
dotted circle $x^{2}+y^{2}=t^{2}$ with radius $r=t=k(1-\varepsilon ).$

The image-point C, $(k\varepsilon ,1,k),$ is on the dotted circle $%
x^{2}+y^{2}=t^{2}$ with the radius $k.$ The image-point D,$k(\frac{\sqrt{2}}{%
2}+\varepsilon ),\frac{\sqrt{2}}{2},k(1+\frac{\sqrt{2}}{2}\varepsilon ),$ is
on the dotted circle $x^{2}+y^{2}=t^{2}$ with radius $r=t=$ $k(1+\frac{\sqrt{%
2}}{2}\varepsilon )$ $\ etc...$

\ The images of the points, contrary to what one might expect, are not
situated on one circular wavefront but, \textit{given the invariance of the
quadratic form} (the dotted circles $x^{2}+y^{2}=t^{2}$), in a circular ring
between $k(1-\varepsilon )\leq r\leq k(1+\varepsilon ),$ \textbf{figure1}).

Let us show now that Poincar\'{e} is right and all the image-points of the
circular wavefront in K' are on an elliptical wavefront. By introducing , in
the system K', the angle $\theta ^{\prime }$ determined by both the radius
vector \textbf{r'} and the Ox' axes, we have $x^{\prime }=r^{\prime }\cos
\theta ^{\prime }$ et $y^{\prime }=r^{\prime }\sin \theta ^{\prime }$. So
with $\ r^{\prime }=t^{\prime }\neq 1$ we have:

\begin{equation}
t=kt^{\prime }(1+\varepsilon cos\theta ^{\prime })
\end{equation}

which is the temporal LT (5), $t=k(t^{\prime }+$ $\varepsilon x^{\prime }),$
with$\ x^{\prime }=r^{\prime }cos\theta ^{\prime }=t^{\prime }cos\theta
^{\prime }.$ We can also write $(r=t)$ the locus of the images-points: 
\begin{equation}
r=kr^{\prime }(1+\varepsilon cos\theta ^{\prime })
\end{equation}

If $r^{\prime }=t^{\prime }=1$ (\textbf{figure1}), we then have:

\begin{equation}
t=r=k(1+\varepsilon cos\theta ^{\prime })
\end{equation}

We will show now that this ''temporal equation'' (6) or ''normal
equation''(7) is the equation of an ellipse in polar coordinates if we
define the polar angle $\theta $ (see \textbf{figure 2} in annex) as the
relativistic transformation of the angle $\theta ^{\prime }$(paragraph 3).

\section{\protect\bigskip Poincar\'{e}'s Elongated Ellipse and the
Relativity of Simultaneity}

Let us first determine Poincar\'{e}'s elongated ellipse in Cartesian
coordinates. We are seeking for the space shape of the wavefront $t^{\prime
}=1$ in K, given the invariance of the quadratic form:

\begin{equation}
x^{2}+y^{2}=t^{2}
\end{equation}
If the time t were \textit{fixed} (see paragraph 4 on Einstein's
synchronisation), we would obviously have a circular wavefront; but $t$
depends by LT on $x^{\prime }$.\textbf{\ }If t' is written in function of
x', we would not have the image of the wave in K. We must write \textit{t in
function of x}\textbf{. }By using the first and the third (x and t) LT (5),
we have respectively if $r^{\prime }=t^{\prime }\neq 1$ $and$ $r^{\prime
}=t^{\prime }=1$: 
\begin{equation}
t=k^{-1}t^{\prime }+\varepsilon x\qquad \qquad t=k^{-1}+\varepsilon x
\end{equation}

We immediately obtain the Cartesian equation of Poincar\'{e}'s elongated
ellipse respectively if $r^{\prime }=t^{\prime }\neq 1$ $and$ $r^{\prime
}=t^{\prime }=1$ (\textbf{figure2)}$:$

\begin{equation}
x^{2}+y^{2}=(k^{-1}t^{\prime }+\varepsilon x)^{2}\qquad \qquad
x^{2}+y^{2}=(k^{-1}+\varepsilon x)^{2}
\end{equation}

%\FRAME{fhFU}{308.75pt}{281.9375pt}{0pt}{\Qcb{Poincar\'{e}'s elongated
%elliptical wavefront and the relativity of simultaneity $(t^{\prime
%}=r^{\prime }=1,$ $\ t^{+}=r^{+}\neq $ $t^{-}=r^{-}).$}}{}{Figure }{\special%
%{language "Scientific Word";type "GRAPHIC";display "USEDEF";valid_file
%"T";width 308.75pt;height 281.9375pt;depth 0pt;original-width
%766.9375pt;original-height 666.4375pt;cropleft "0";croptop "1";cropright
%"1";cropbottom "0";tempfilename 'I6795Y01.wmf';tempfile-properties "XPR";}}

At once we check that Poincar\'{e}'s ellipse, by replacing x'

\begin{equation}
x^{\prime }=k^{-1}x-\varepsilon t^{\prime }
\end{equation}
in $x^{\prime 2}+y^{\prime 2}=t^{\prime 2}=1_{t^{\prime }}$ (4) ,
respectively if $r^{\prime }=t^{\prime }\neq 1$ $and$ $r^{\prime }=t^{\prime
}=1$, can be also written thus$:$

\begin{equation}
(k^{-1}x-\varepsilon t^{\prime })^{2}+y^{2}=t^{\prime 2}\qquad \qquad
(k^{-1}x-\varepsilon )^{2}+y^{2}=1_{t^{\prime }}
\end{equation}
The image-points (\textbf{figure 2}) are\ situated on Poincar\'{e}'s
elongated ellipse\footnote{%
It is the inverse case that is explicitely considered by Poincar\'{e}
(historical introduction).}, with Observer O at the focus F and Source S at
the centre C. The eccentricity of the ellipse is $\varepsilon =\frac{%
k\varepsilon }{k}$ where k is the length of the great axis (we choose, in
figure 2, the small axis of the ellipse $r^{\prime }=t^{\prime }=1$). The
equation of Poincar\'{e}'s ellipse can be written in polar coordinates with
pole $O$, focus $F$ and the polar angle $\theta $ defined in K (with both
standard parameters of the ellipse $e,p$): 
\begin{equation}
r=\frac{p}{1-e\cos \theta }
\end{equation}

with the small axe of the ellipse $b=r^{\prime }=1$

\begin{equation*}
p=a(1-\varepsilon ^{2})=ak^{-2}=kk^{-2}=k^{-1}
\end{equation*}

we immediately deduce the polar equation of Poincar\'{e}'s ellipse 
\begin{equation}
r=\frac{\sqrt{1-\varepsilon ^{2}}}{1-\varepsilon \cos \theta }=\frac{1}{%
k(1-\varepsilon \cos \theta )}
\end{equation}

with eccentricity $e=\frac{f}{a}=\frac{k\varepsilon }{k}=\varepsilon $ and
with the two standard parameters of the special relativity $\varepsilon ,$ $%
k:$

\begin{equation*}
a^{2}-f^{2}=b^{2}\qquad \qquad k^{2}-\varepsilon ^{2}k^{2}=1
\end{equation*}

If $r^{\prime }=t^{\prime }\neq 1$, we have the equation of the ellipse

\begin{equation}
r=\frac{\sqrt{1-\varepsilon ^{2}}}{1-\varepsilon \cos \theta }=\frac{%
r^{\prime }}{k(1-\varepsilon \cos \theta )}
\end{equation}
with $r^{\prime 2}(k^{2}-\varepsilon ^{2}k^{2})=r^{\prime 2}.$

It should be reminded that the ''normal equation'' (7) of the ellipse is

\begin{equation}
r=kr^{\prime }(1+\varepsilon cos\theta ^{\prime })
\end{equation}

\bigskip Thus we obtain from (16 and 7) the formula of relativistic
transformation of angle 
\begin{equation}
\cos \theta =\frac{\cos \theta ^{\prime }+\varepsilon }{1+\varepsilon \cos
\theta ^{\prime }}
\end{equation}

So it is now utterly demonstrated that Poincar\'{e} is right and that 
\textit{the geometrical image by LT of a circular wavefront is an elongated
ellipse} its polar equation being (16) and its Cartesian equation being
(12). Poincar\'{e}'s ellipse gives the other formulae of aberration, in
particular:

\begin{equation}
\sin \theta =\frac{\sqrt{1-\varepsilon ^{2}}}{1+\varepsilon \cos \theta
^{\prime }}\sin \theta ^{\prime }
\end{equation}

It is now essential to interpret the historical case (see introduction and
footnote 2) considered by Poincar\'{e} (in connection with Michelson's
experiment where the source is on the Earth, see conclusion): the circular
light wavefronts are developed around O (the ether is now \textit{by
definition} at rest relative to K): $r=t=1$. What is the image of the
circular locus of the points (determined now by $\theta )$ seen from O'
(where the source is at rest in K', ''system of the Earth'')? Given that
Poincar\'{e}'s ellipse I, in the first case, is directly inscribed in LT, it
is easy to define Poincar\'{e}'s ellipse II, in the second case, both by
inverting in (7) the primed and the unprimed and by changing the sign of $%
\varepsilon .$ The ''normal'' equation of Poincar\'{e}'s (historical)
ellipse II is therefore:

\begin{equation}
r^{\prime }=kr(1-\varepsilon \cos \theta )
\end{equation}

\bigskip The polar equation of Poincar\'{e}'s ellipse II, its source S (in
O', ''on the Earth'' ) in moving occupies the focus F$^{\ast }$ (see \textbf{%
figure 2}) and the observer O occupies the centre C is (with 18):

\begin{equation}
r^{\prime }=r\frac{1}{k(1+\varepsilon \cos \theta ^{\prime })}
\end{equation}

The ''normal'' (temporal) equation of the ellipse I then is the polar
equation of the ellipse II. The Cartesian equation of Poincar\'{e}'s ellipse
II is:\qquad 
\begin{equation}
(k^{-1}x^{\prime }+\varepsilon t)^{2}+y^{\prime 2}=t^{2}
\end{equation}

So in Poincar\'{e}'s relativistic kinematics we can have, with no
contradiction at all, an \textit{elliptical wavefront} \textit{in the system
of the source}.

What is now the \textbf{physical interpretation} of Poincar\'{e}'s elongated
ellipse? We underline, at this stage, three points (for the relativistic
Doppler effect, see conclusion):

\bigskip \textbf{1)} Poincar\'{e}'s elongated ellipse is the direct
translation of the ''\textbf{headlight effect}'': the isotropic emission of
a moving source is not isotropic seen from the rest system (relativistic
transformation of angles $\theta ^{\prime }$into $\theta $). In three
dimensions of space the reduction of the angle of aperture of the cone of
emission of a moving source is physically (synchrotron radiation,
bremsstrahlung...) represented, on the whole (from any angle), by the
ellipsoidal shape of the wavefront.

\textbf{2)} Poincar\'{e}'s elongated ellipse is the direct translation of
the \textbf{relativity of simultaneity}: the set of simultaneous events in
K' of the spherical wavefront in time t' is \textit{not} a set of
simultaneous events in K, in time t. In particular, if the two events $%
(1,0,1 $) et $(-1,0,1$) are simultaneous in K', they are \textit{not}
simultaneous (6), $k(1+\varepsilon ),0,k(1+\varepsilon )$ and $%
k(1-\varepsilon ),0,k(1-\varepsilon ),$ in K. Let us also note that the
image of the distance $"2"$ between these two events in K' is elongated $"2k"
$ in K. These two fundamental points put the emphasis on the fact that
Poincar\'{e}'s ellipse is not only a geometrical \textit{image} but also a
physical \textit{shape} of the wavefront.

\textbf{3)} Poincar\'{e}'s elongated ellipse is the direct translation of
Poincar\'{e}'s completely relativistic\footnote{%
Poincar\'{e} is philosophically very anti-absolutist as well \cite{13}.}
ether: \textit{put the ether at rest in one (K) or in the other system (K'})
is exactly equivalent to \textit{define the ellipse with the direct LT or
the inverse LT}. So in Poincar\'{e}'s own words: if t' is the true time \
(''circular'' time), t then is the local time (''elliptical'' time) and
inversely (by LT) if t is the true time (''circular'' time), t' then is the
local time (''elliptical'' time). That is completely relativistic and
Poincar\'{e}'s elongated ellipse is ''the immediate interpretation of
Michelson experimental result''(see conclusion).

Poincar\'{e}'s ether is relativistic but \textit{not deleted} (as Einstein's
one) because it remains \textit{the relativistic definition of state of rest}%
: when we choose by definition ether at rest in one system (spheres or 
\textit{true time}), it is not at rest in the other system (ellipsoids or 
\textit{local time}).

Objectively we have two possibilities to choose the criterion of the \textit{%
relativistic state of rest of a system}:\textbf{\ the source of light or the
medium of light}. In Einstein-Minkowski's relativistic \textit{kinematics},
the criterion is clearly the \textbf{source} (\textit{the proper system, see
paragraph 4}). In Poincar\'{e}'s\ relativistic \textit{kinematics}, the
criterion is clearly the \textbf{ether (''circular waves'')}.

That is a paramount difference because in \textit{Einstein-Minkowski's
proper system (see paragraph 4)} we always have by definition spherical
waves or in other words, \textit{the equality between forth travel time and
back travel time}. It is not the case with Poincar\'{e}'s definition of
units where we can have without any contradiction, an elliptical wavefront
(a local time) \textit{in the system of the source} (see conclusion).
Poincar\'{e}'s relativistic duality between true time and local time \textit{%
doesn't correspond} to Einstein-Minkowski's relativistic duality between
proper time and improper time (paragraph 4).

\section{Einstein's Kinematics: identical Spheres, identical rigid Rods and
Convention of Synchronisation}

If according to Einstein, the object (1) and the image (2), are both
spherical and \textit{concentric} within the two systems, then\ two
simultaneous events in K, for example (1, 0, 1) and (-1, 0, 1), must be also
simultaneous in k.

That seems in contradiction not only with Poincar\'{e}'s ellipse but also
with Einstein's well known definition of relativity of simultaneity.\
Therefore the image by LT in k of a spherical wave in K \textbf{cannot be a
spherical wave}. So it could appear at this stage that \textit{Poincar\'{e}
is right} and \textit{Einstein is wrong}.

However the question is: ''What in Einstein's reasoning is true?''. Let us
return to Einstein's 1905 quotation (paragraph 1). The two quadratic forms $%
x^{2}+y^{2}+z^{2}=c^{2}t^{2}$ and $\xi ^{2}+\eta ^{2}+\zeta ^{2}=c^{2}\tau
^{2}$ are the \textit{geometrical} equations of two spheres (two circles in
two dimensions or two equidistant points respectively from O and O' in one
dimension). Let us note that the young Einstein doesn't specify, in the
previous quotation, in \textit{which system} the source is at rest\textit{. }
So if we consider\footnote{%
We can also consider that the emission is an \textbf{event} in a strong
Einstein's meaning and an event has no velocity. We find in Einstein's
introduction the enigmatic sentence '' The introduction of a ''lumineferous
ether'' will prove to be superfluous inasmuch as the view here to be
developped will not require an ''absolutely stationary space provided with
special properties \textit{nor assign a velocity-vector to a point of the
empty space in which electromagnetic processes take place}.'' The emission
by a source of light is an event that \textit{has no velocity} (see note 7)
and therefore everything happens as if the source were at rest within each
system). \cite[concept d'\'ev\'enement]{8}} now that we have \textit{two
identical sources,} in O' and O, emitting a signal of light \textit{%
simultaneously} at the time, 
\begin{equation}
\tau =t=0
\end{equation}
the physical situation is perfectly identical in each system (Einstein's
deletion of ether\footnote{%
Einstein's deletion of ether is completely inseparable of Einstein's photon
(1905)\cite{7}.}). So we must have two identical spherical wavefronts, $%
x^{2}+y^{2}+z^{2}=c^{2}$ around O\ and $\xi ^{2}+\eta ^{2}+\varsigma
^{2}=c^{2}$ around O', simultaneously at the time 
\begin{equation}
\tau =t=1_{t}=1_{\tau }
\end{equation}
It immediately follows from the latter choice of two \textit{identical }%
units of time that we have two identical units of length $1_{x}$ $=$ $1_{\xi
}=c1_{t}=c1_{\tau }$. Let us point out that the travel time of the circular
wavefront either to the right or to the left (on the $x,$ $\xi $ axis) are
identical within the two systems. Einstein's definition of identical units
of time is therefore completely coherent with Einstein's\textbf{\ identical
rigid rods}\textit{\ (}$1_{t}=1_{\tau }$ is on the other hand incompatible
with Poincar\'{e}'s definition of units, see conclusion). Einstein writes in
this sense in 1905:

\begin{quotation}
Let there be given a stationary rigid rod; and let its length be $L_{0}$ as
measured by a measuring-rod which is also stationary. \ In accordance with
the principle of relativity \textit{the length of the rod in the moving
system}\ - must be equal to \textit{the length }$L_{0}$ \textit{of the
stationary rod }. \cite{2}
\end{quotation}

In this respect, M. Born is perhaps the only physicist who underlined that
the young Einstein introduces in fact a tacit assumption (1921):

\begin{quotation}
A fixed rod that is at rest in the system K and is of length 1 cm, will, of
course, also have the length 1 cm, when it is at rest in the system k. We
may call this \textbf{tacit assumption} of Einstein's theory the \textit{%
principle of the physical identity of the units of measure}. \ \cite{1}
\end{quotation}

Einstein's principle of identity (see also \ \cite{21}) stipulates $%
L_{0}=1_{x}$ $=$ $1_{\xi }=c1_{t}=c1_{\tau }$ and therefore (24) becomes

\begin{equation}
\tau =t=\frac{L_{0}}{c}
\end{equation}

There are two Einstein's spherical waves and each spherical wave\ defines,
in one dimension, two simultaneous events, (-1, 0, 1) and (-1, 0, 1), within
each system (K and k). In other words: \textbf{Einstein's rigid rod }$%
\mathbf{(2}L_{0})$\textbf{\ is defined by two simultaneous events within
each system (K and k)}. Einstein's spherical waves are not in contradiction
with ''Einstein's relativity of simultaneity (after LT)'' because it is
''Einstein's convention of simultaneity (before LT)'' or in other words
''Einstein's \textit{convention of synchronisation of identical clocks in A\
and B} with the exchange of a signal of light in K (before LT)''. Let us now
demonstrate that point by rigorously distinguishing the two stages of
Einstein's deduction: before LT and after LT.

\subsection{Before LT (the proper Systems)}

According to the young Einstein, ''It is essential to have time defined by
means of \textit{stationary} clocks in \textit{stationary }system''.
Einstein's famous repetition of the concept ''stationary\footnote{%
The ether is at rest (or stationary) within the two systems and it is
therefore superfluous. The deletion of the medium of light involves the
relativistic state of rest is defined, by Einstein's repetition with respect
to the source of light.}'' is essential because he notices about his second
system k ($\xi $, $\eta $, $\zeta $, $\tau $):

\begin{quotation}
To do this [deduce LT] we have to express in equations that $\tau $ is
nothing else than the set of data of clocks at rest in system k $\ $, which
have been synchronized [$A^{\prime }B^{\prime }$ ] according to the rule
given in paragraph 1 [$AB$ ] \ \cite{2}.
\end{quotation}

Without any loss of generality we make $A\equiv O$ (respectively $A^{\prime
}\equiv O^{\prime })$ in young Einstein's notations (and then $t_{A}=\tau
_{A^{\prime }}=0$). We have $2t_{B}=t_{O}^{\ast }$ in K (respectively $2\tau
_{B^{\prime }}=\tau _{O^{\prime }}^{\ast }$ in k ) and $c=2\frac{OB}{%
t_{O}^{\ast }}$ in K (respectively $c=2\frac{O^{\prime }B^{\prime }}{\tau
_{O^{\prime }}^{\ast }}$ in k$),$ with $L_{0}=OB=O^{\prime }B^{\prime },$
where $t_{O}=0$ (respectively $\tau _{O^{\prime }}=0$) is the time of
emission of the light signal in K (respectively in k) and $t_{O}^{\ast }$ \
(respectively $\tau _{O^{\prime }}^{\ast })$ is the time of reception of the
light signal in B in K (respectively in B' k). \ 
\begin{equation}
t_{O}^{\ast }=\tau _{O^{\prime }}^{\ast }=2\frac{L_{0}}{c}=T_{0}
\end{equation}
Given that forth travel time and back\ travel time are identical with 
\begin{equation}
t_{O}=\tau _{O^{\prime }}=0
\end{equation}
we finally have

\begin{equation}
t_{B}=\tau _{B^{\prime }}=\frac{L_{0}}{c}=\frac{1}{2}T_{0}.
\end{equation}

Einstein's interpretation of the invariant quadratic form as \textbf{two
physical spherical wavefronts} (1 \& 2) is therefore exactly the same
concept (see equations 23 and 27, 25 and 28) as Einstein's 1905 \textbf{%
convention of synchronisation within the two systems} (\textit{in one
dimension} where forth wavefront becomes forth time travel and back
wavefront becomes back time travel). The proper time T$_{0}$(index ''zero''
means ''proper''), the duration between \textit{two events at the same place}%
, in young Einstein's notation is $t_{A}^{\ast }$ \ (respectively $\tau
_{A^{\prime }}^{\ast })$ or $t_{O}^{\ast }$ (respectively $\tau _{O^{\prime
}}^{\ast }$) with the identity between forth travel time and back travel
time (factor $\frac{1}{2})$. This is Einstein's synchronisation \textit{%
without contraction} :$OB=O^{\prime }B^{\prime }$ (For Poincar\'{e}'s
convention of synchronisation with contraction see \cite{17} and \cite{7}).
We point out that Einstein's convention is not based on the choice of only
one unit of length in one system (see Poincar\'{e}, conclusion) but on two
identical units of length (B' is not the image by LT of B) within each
system.

Historically Einstein deduced the identical units of time T$_{0}$ from the
identical units of length L$_{0}$  and from invariance of ''one way speed of
light'' (see forth travel time and back travel time or the circular waves).
We can as well, like Minkowski \cite{5}, reverse that situation by defining
first a proper time $T_{0}$ \ and next the proper length L$_{0}$. This
inversion seems avoid the rigid rods but in fact nothing is changed because
Einstein's two spherical waves, Einstein's convention of synchronisation,
Einstein's one way speed of light \cite{19}, Einstein's rigid rods L$_{0}$
and Minkowski's ''Eigenzeit'' are completely inseparable. Without Einstein's
isotropy (spheres 1 \& 2), there is no Minkowski's Eigenzeit because the
relativistic state of rest is defined relatively to the source (see
paragraph 3, third point and conclusion): ''identical units within each
system'' and ''isotropy in each proper system'' are exactly the same concept.

The main result in the framework of this paper concerns Einstein's ''one way
speed of ligth'' definition of proper length $L_{0}$ with half of proper
time, $\frac{1}{2}T_{0}$. What does it mean? Einstein defines first the
simultaneity of two events at the same place ($A=B$). Secondly he defines
the simultaneity of two events at different places $(A\neq B)$. But the
departure (from A), the arrival (in B) and the return (in A) of the light
are 3 \textit{successive} events. What are finally Einstein's two
simultaneous events in A\ and B? These two events are in k: ($\xi _{A},\frac{%
1}{2}T_{0})$ and $(\xi _{B},\frac{1}{2}T_{0}).$ These are the two ends of
the rigid rod at the same time. So we have Einstein's \textit{relativistic}
(one way) definition of rigidity::

\begin{definition}
The \textbf{proper length, }$L_{0}=$ $\xi _{B}-\xi _{A},$ is defined \textbf{%
by two events at the same time,} $\tau =\frac{1}{2}T_{0},$ in the proper
system k (respectively, \textbf{\ }$L_{0}=$ $x_{2}-x_{1},$ \textbf{at the
same time,} $t=\frac{1}{2}T_{0}$, in the proper system K).
\end{definition}

\subsection{After LT (the improper Systems)}

\bigskip\ Einstein's construction of invariant quadratic form as physical
spherical wavefront means that Einstein (and Minkowski) defines the units
before taking in account the relative velocity v $:$ Einstein's units are\
completely \textit{independent }\footnote{%
Einstein's spheres are not necessarily identical. Fock V. underlined about
the scale factor: ''\textit{We have }$c^{2}\tau ^{2}-\xi ^{2}-\eta
^{2}-\zeta ^{2}=\varphi ^{2}(x,y,z,t)$\textit{\ }$(c^{2}t^{2}-$\textit{\ }$%
x^{2}-y^{2}-z^{2}$\textit{\ ). The factor }$\varphi ^{2}$\textit{, or rather 
}$\varphi $\textit{, evidently characterises the ratio of the scales of
measurement in the primed and unprimed frames. Further, it follows that this
factor cannot depend on the relative velocity. It is usually said, following
Einstein, that the scale factor can ''evidently'' depend on nothing but the
relative velocity, and it is subsequently proved that, in fact, it does not
have any dependence but is equal to 1: }$\varphi (x,y,z,t)=1$\textit{.''} 
\cite[principe d'identité]{9}. In Poincar\'{e}'s relativistic kinematics the
dependance of the scale factor on the velocity, $l(\varepsilon ),$ is
paramount. The group property of LT proves that the velocity is a relative
velocity.} on the relative velocity v (and also $\beta $ and $\gamma )$.
Therefore there is no contradiction with LT, because the definition of
space-time units is a preparation of the two systems not only prior to the
use of LT and even, more fundamentally, prior to the\textit{\ deduction of LT%
} (see note 4). Einstein's 1905 deduction of LT is very complicated but
Einstein's immediate 1907 deduction of LT from the invariance of the
quadratic form, i.e. the invariance of (one way \cite{19}) speed of light,
has become a classic \cite{3}:

\begin{equation}
\xi =\gamma (x-vt)\qquad \qquad \eta =y\qquad \qquad \tau =\gamma (t-\frac{v%
}{c^{2}}x)
\end{equation}

If Einstein's definition of space-time units is not in contradiction with
LT, it requires on the other hand a specific use of LT. In young Einstein's
words: ''We will call the length to be discovered $L$ \textit{the length of
the (moving) rod in the stationary system''\cite{2}. }In current words, the
length to be discover by LT is the improper length L, and also the improper
time T, respectively relatively to the proper length L$_{0}$ or the proper
time $T_{0}.$ The role of the LT consists fundamentally of introducing the
velocity v or defining the \textit{improper moving} system (k relative to
proper K or inversely). In all standard books we can find Einstein's
deduction, with the use of LT, of the dilation of proper time $T=\gamma
T_{0} $ and the \textbf{contraction }of proper length $\gamma ^{-1}L_{0}.$
Therefore the\textit{\ improper time and the improper length (in the moving
system) are \textbf{inversely }}proportional\textit{. }Let us examine
Einstein's use of LT (the standard deduction) in details.

\subsubsection{\textbf{DILATION OF PROPER TIME}}

\ The proper time, $T_{0}=\tau _{2}-\tau _{1},$ is the duration between this
two events \textit{at the same place} ($\xi _{1}$ = $\xi _{2}=\xi )$ in k.
We find the duration T in K by the \textit{second LT:}

\begin{equation*}
t_{1}=\gamma (\tau _{1}-\frac{v}{c^{2}}\xi )\qquad \qquad \qquad
t_{2}=\gamma (\tau _{2}-\frac{v}{c^{2}}\xi )
\end{equation*}

The duration, $T=t_{2}-t_{1}$, in the moving system K is 
\begin{equation}
T=\gamma T_{0}
\end{equation}

With the first LT we remark that the two considered events are not at the
same place in K

\begin{equation}
x_{1}=\gamma (\xi -v\tau _{1})\qquad \qquad \qquad x_{2}=\gamma (\xi -v\tau
_{2})
\end{equation}

\bigskip This is a very well known result: Einstein(-Poincar\'{e}'s, see
conclusion) dilation is the consequence of the fact that we must use \textit{%
two clocks in different places} ($\Delta x=vT_{0}$ ) of the moving system.

\subsubsection{\textbf{CONTRACTION OF PROPER LENGTH}}

According to Einstein (as in all standard books on SR), the proper length $%
L_{0}=$ $\xi _{2}-\xi _{1}$ is the length at rest in k $(\xi _{2}$, $\xi _{1}
$ are the coordinates of the ends of the rod in k). The length of the moving
rod is then defined as the distance between the two ends of the rod \textit{%
at the same time} ($t=t_{1}=t_{2})$ in K. We immediately find this length, $%
L=$ $x_{2}-x_{1}$, by the inverse first LT:

\begin{equation*}
\xi _{1}=\gamma (x_{1}+vt)\qquad \qquad \qquad \xi _{2}=\gamma (x_{2}+vt)
\end{equation*}

and therefore we obtain Einstein's famous contraction:

\begin{equation}
L=\gamma ^{-1}L_{0}
\end{equation}

\bigskip Both Einstein's deductions, dilation of time and contraction of
length, are presented in all standards books as perfectly symmetric: two 
\textit{events} at the same place (in k) for the dilation of duration and
two \textit{events} at the same time (in K) for the contraction length.
Nevertheless: what are the complete coordinates of the two events (ends of
the rods) in the proper system k? In order to have the complete symmetry, we
must consider the other LT not only in the case of dilation of duration (31)
but also in the case of contraction of length. The second LT is:

\begin{equation}
\tau _{1}=\gamma (t+\frac{v}{c^{2}}x_{1})\qquad \qquad \qquad \tau
_{2}=\gamma (t+\frac{v}{c^{2}}x_{2})
\end{equation}

This is a completely ignored result. The second LT determines obviously the
times, $\tau _{1}$ $and$ $\tau _{2}$, of the ends of the rods $\xi _{1}$and $%
\xi _{2}$ in the proper system k and thus the complete coordinates of the
two events ($\xi _{1,}$ $\tau _{1})$ and ($\xi _{2}$,$\tau _{2})$: the
simultaneous events in k are, obviously \textbf{by LT}, not simultaneous
events ($\Delta \tau =\frac{v}{c^{2}}L_{0}$ ) in k. This is in contradiction
with Einstein's definition of identical RIGID rods (see above \textbf{%
Definition 1} in 4-1) that implies that the proper length must be defined by
simultaneous events in the proper system (the ends of the rigid rods are
defined at the same time $\tau $). So \textbf{Einstein's contraction} is 
\textbf{not }deduced directly from LT: \textit{it is a supplementary
hypotheses }(this is not the case in Poincar\'{e}'s kinematics, see
conclusion).

\begin{definition}
The proper length $L_{0}$ is defined by two simultaneous events ($\xi _{1,}$ 
$\tau )$ and ($\xi _{2}$, $\tau )$ in k (definition 1) and the improper
length is defined by two simultaneous events ($x_{1,}$ $t)$ and ($x_{2}$, $t)
$ in K. But these events are \textit{not} the images by LT one another.%
\textit{\ }
\end{definition}

\textit{Einstein's inverse }$(\gamma ^{-1})$ c\textit{ontraction (32) or
''Einstein's breaking of symmetry'' is therefore clearly in opposition with
Poincar\'{e}'s \textbf{direct} proportionality of the transformation of time 
\textbf{and length} in the moving system (see conclusion)}

\section{Conclusion: Definition of space-time Units in \newline
Poincar\'{e}'s Relativistic Kinematics.}

\bigskip Poincar\'{e} writes in 1911 in ''L'espace et le temps'' on the
special theory of relativity:

\begin{quotation}
Today some physicists want to adopt a new convention. This is not that they
have to do it; they consider that this convention is easier, that's all; and
those who have another opinion may legitimately keep the old assumption in
order not to disturb their old habits.\cite{16}
\end{quotation}

\bigskip ''Some physicists'' is a clear allusion to Einstein and Minkowski.
What is the difference, according to Poincar\'{e}, between the ''old
convention'' and the ''new convention''? Let us examine Poincar\'{e}'s old
(tacit) assumption in detail. What happens if we place another source in the
second system K in Poincar\'{e}'s relativistic kinematics? Suppose that the
relativistic ether is by definition at rest (spheres around O') relative to
the first source in K'. Poincar\'{e}'s relativistic ether is then moving
relative to the second source at O in K and so we rediscover the second case
with an ellipsoidal wave in the system of the source. And reciprocally, with
inverse LT, the role of the ether (\textit{the criterion of relativistic rest%
}) is inverted. Logically in Poincar\'{e}'s SR, with one source or two
sources, we \textbf{always} have a \textit{sphere} in one system and an%
\textit{\ ellipsoid} in the other system and \textbf{never }two spheres in
the two systems (paragraph 4).

\ If historically Poincar\'{e} deduced directly the ellipsoid from the
contraction of unit, we must now deduce the contraction of unit from the
ellipsoid directly provided by the LT. From the main property of an
elongated ellipse $r^{+}+r^{-}=2kr^{\prime }$ (see \textbf{figure 2} the
forth distance $r^{+}$ and the back distance $r^{-}$ with respect to the
second focus F$^{\ast }$ or the forth travel time t$^{+}$ and the back
travel time t$^{-}$ with respect to the second focus $F^{\ast }$)\footnote{%
Poincar\'{e}'s \textit{exact }synchronisation (at the second order) is
developped with Poincar\'{e}'s elongated ellipsoid in \cite{7} (1999).
Poincar\'{e}'s elongated ellipse ($t^+ \neq t^-$) \textit{is }
Poincar\'{e}'s convention of synchronisation.} where M means ''mean
(average)'', ''round trip'' or ''two ways'', we obtain:

\begin{equation}
r_{M}=\frac{r^{+}+r^{-}}{2}=kr^{\prime }\qquad \qquad t_{M}=\frac{t^{+}+t^{-}%
}{2}=kt^{\prime }
\end{equation}

We have by definition in the system K' $t^{\prime }=1_{t^{\prime }\text{ }%
}and$ $r^{\prime }=1_{r^{\prime }}$ (the choice of only one length unit). So
if the elongated ellipse is an alternative definition of the units we must
be able to deduce immediately Poincar\'{e}'s \textit{''round trip'' units}
in K. Indeed we have:

\begin{equation}
1_{r}=k1_{r^{\prime }}\qquad \qquad \qquad 1_{t}=k1_{t^{\prime }}
\end{equation}

The unit of local time (''elliptical time'') $1_{t}$ is always \textit{%
dilated }in relation with the unit of true time (''circular time'') $%
1_{t^{\prime }}.$

\ For the unit of space, we must first show that there is no transversal
contraction :

\begin{equation*}
1_{y}=1_{r}\sin \theta \text{ et }1_{y^{\prime }}=1_{r^{\prime }}\sin \theta
^{\prime }
\end{equation*}

with $\theta ^{\prime }=\frac{\pi }{2}$, we have (19) $sin$ $\theta =k^{-1}$
and thus 
\begin{equation}
1_{y}=1_{y^{\prime }}
\end{equation}
We immediately have for the longitudinal component $\cos \theta =\cos \theta
^{\prime }=1$, with $r^{+}=k(1+\varepsilon )$ and $r^{-}=k(1-\varepsilon )$

\begin{equation}
1_{x}=k1_{x^{\prime }}
\end{equation}

Let us call $1_{x^{\prime }}$ ''the unit at (relativistic) rest'' and $1_{x}$
''the unit in (relativistic) moving''. So the unit at rest $1_{x^{\prime }}$
is seen\ purely longitudinally elongated (by a factor k$)$ by the observer O
in moving in K. This is an unusual language but if we inverse the situation 
%(if A\TEXTsymbol{>}B $\Longrightarrow $ B\TEXTsymbol{<}A)
(if $A > B$ $\Longrightarrow$ $B < A$) we then have

\begin{equation}
1_{x^{\prime }}=k^{-1}1_{x}
\end{equation}

The unit 1$_{x}$ in moving is seen longitudinally contracted (by a factor k$%
^{-1})$ from the observer at rest O'. This is a more usual language\footnote{%
In Einstein's relativistic logic %A\TEXTsymbol{>}B
$A>B$ has no meaning because there are identical rigids rods (A=B) within
the two systems (see definitions 1 and 2 in paragraph 4). In Poincar\'{e}'s
relativistic logic it is completely equivalent to say ''the image of the
unit at rest is always elongated'' and ''the image of an unit in moving is
always contracted''.}. We rediscover therefore the initial postulate of
Poincar\'{e} about the contraction of a moving unit (see historical
introduction).

We underline that Poincar\'{e}'s deduction of \textit{dilated units} is
based, and \textit{only }based, on the application of LT (the ''old
convention''). He doesn't need like Einstein a supplementary hypotheses (see
paragraph 4). Einstein-Minkowski's definition of identical units within both
systems is clearly \textit{beyond} the LT (the ''new convention'', see
paragraph 4). At the end of the deduction of his elongated ellipse
Poincar\'{e} writes:

\begin{quotation}
This hypothesis of Lorentz and FitzGerald will appear most extraordinary at
first sight. All that can be said in its favour for the moment is that it is
merely the immediate interpretation of Michelson experimental result, if we 
\textit{define} (in italics in the text) distances by the time taken by
light to traverse them.\cite{14}
\end{quotation}

So with the ellipse we see immediately that the time of the \textit{round
trip} is the same in all directions (and therefore for the two Michelson's
perpendicular directions). So Poincar\'{e}'s historical ellipse is the
immediate interpretation of Michelson's null result\footnote{%
And also an immediate explanation of Sagnac \textit{non null} result (1913).
According to Selleri one of the main problems of rotating platform with
Einstein's kinematics is precisely Einstein's invariance of one way speed of
light, $t^{+}=t^{-},$ in the proper system \cite{19}. In Poincar\'{e}'s
relativistic kinematics we can have in the system of the source $t^{+}\neq
t^{-}$ (see figure 2)$.$ With Poincar\'{e}'s elongated ellipse \cite{7} and
Poincar\'{e}'s group with rotations \cite{17}, we predict immediately (at
the second order k) the experimentally measured difference of time $\frac{%
t^{+}-t^{-}}{2}=k\varepsilon L$ ($L=2\pi R$, R being the radius of the
platform).} without postulating that the source in the system (proper) of
the Earth emits spherical waves.

And this is not all: according to Poincar\'{e} the distances are defined by
the dilated time taken by light to traverse them. We can deduce this
fundamental point directly from LT. The\ usual definition of the length of a
rod implies that we consider \textit{at the same time} the two ends of the
rod. So we consider the two ends of the unit of length $1_{x^{\prime }}$ in
K' at the same time $t^{\prime }=0$ (the primed coordinates are $0,0$ and $%
1,0$). What is the length of the rod in the other system K (the moving
system) according to Poincar\'{e}, i.e. according to LT? The calculation
with (5) gives immediately $1_{x}=k1_{x^{\prime }}.$ The elongation in the
moving system of the stationary rod is a direct consequence of the fact that
two simultaneous events in K' are not simultaneous events in K (see
paragraph 4) .

We conclude by remarking that Poincar\'{e}'s relativistic kinematics is
based on a fundamental \textit{space-time proportionality} (a dilation by a
factor k) in perfect harmony with the invariance of the speed of light.

\begin{equation}
\frac{r_{M}}{t_{M}}=\frac{kr^{\prime }}{kt^{\prime }}=\frac{1_{r^{\prime }}}{%
1_{t^{\prime }}}=\frac{k1_{r^{\prime }}}{k1_{t^{\prime }}}=c=1
\end{equation}

Poincar\'{e}'s \textbf{direct space-time proportionality (35 \&37)} (very
strange in Einstein's kinematics\footnote{%
In Einstein's kinematics, the image by (30 \& 32 ) of a purely longitudinal
light clock in the proper system implies that the velocity of light in the
moving system is $\gamma ^{-2}$ $c$. This is perhaps the reason why we 
\textit{only} find, in Einstein's texts the purely transversal light clocks
(without contraction) !}, see paragraph 4, (\textbf{30 \& 32}))
characterizes fundamental Poincar\'{e}'s choice of space-time units in
relativistic kinematics \textit{I shall choose the units of length and of
time in such a way that the velocity of light is equal to unity }($\lambda
^{\prime }\nu ^{\prime }=\lambda \nu =c=1)$\textit{. }This is the reason why
we kept Poincar\'{e}'s notations $\varepsilon $ ($\beta )$ and $k$ $(\gamma )
$: behind Poincar\'{e}'s notations, there is not only Poincar\'{e}'s
perfectly symmetrical representation of LT (5) but also Poincar\'{e}'s ''old
convention'' about the \textbf{metric }(in the sense of space-time units of
measure) underlying the invariance of quadratic form in SR.

The existence of a ''fine structure'' of SR (two very close but not merged
theories) is therefore demonstrated \cite{7}.

According to Minkowski (1908), Einstein's SR was not a local theory but a
theory of ''the world'' or '' the Universe'' (worldline, worldpoint,
worldinterval and even world principle, which is Minkowski's name for the
principle of relativity \cite{5}).

The problem is that Minkowski's metric is incompatible with an expansion of
the Universe (Hubble 1929). Finally we point out that \textit{Poincar\'{e}'s
metric}\textbf{\ } involves not only a dilation of time but also an \textbf{%
expansion }of space. We will show in another paper that Poincar\'{e}'s\ 
\textit{completely relativistic} expansion is directly connected with the
deduction of the relativistic Doppler formulae from Poincar\'{e}'s ellipse.

\section{Annex: Penrose's and Poincar\'{e}'s elongated ellipsoid}

The question under discussion is directly connected to another question:
Penrose-Terrel's analysis on ''The Apparent Shape of a Relativistic Moving
Sphere'' (1959) or ''The Invisibility of Lorentz Contraction'' (1959) in
Einstein's SR. If we search the apparent shape for \textit{one} observer of
a moving material sphere, according to Penrose \cite{6}, we have to send a
signal of light that is reflected on the surface of the sphere and that
finally returns to the observer. Penrose shows that we have to take into
account Einstein's 1905 relativistic\ formulae of aberration and Doppler
effect. Terrel writes thus:

\begin{quotation}
The factor M is the magnification, the ratio between subtended angles as
seen by the observers O' and O, or the ratio of apparent distances of the
objects from the two observers. It is interesting that M is precisely the
Doppler shift factor becoming $\sqrt{\frac{1-\frac{v}{c}}{1+\frac{v}{c}}}$
for $\theta =0=\theta ^{\prime }.$ \cite{20}
\end{quotation}

In one dimension we avoid the question of aberration (for $\theta =0=\theta
^{\prime }),$ which is the main problem of Penrose-Terrel and not under
discussion in the present paper. If we try to measure a moving contracted
rod $L=\gamma ^{-1}L_{0}$ with the mean time travel of the signal of light
(forth $+$ and back $-$travel), Lampa \cite{4}, before Penrose and Terrel,
shows that, the longitudinal Doppler effect is respectively $\nu ^{+}=\sqrt{%
\frac{1-\frac{v}{c}}{1+\frac{v}{c}}}$ and $\nu ^{-}=\sqrt{\frac{1+\frac{v}{c}%
}{1-\frac{v}{c}}.}$ We have the mean travel time\footnote{%
In the deduction of Poincar\'{e}'s ellipse we have immediately by LT: $%
t^{+}=k(1+\varepsilon )$ and $t^{-}=k(1-\varepsilon )$.} $t^{+}=\sqrt{\frac{%
1+\frac{v}{c}}{1-\frac{v}{c}}}$ $+$ $t^{-}=\sqrt{\frac{1-\frac{v}{c}}{1+%
\frac{v}{c}}=}$ $\gamma T$. The mean ''apparent distance'' from O is,
according to Lampa, $L_{app}=\gamma L.$ Penrose explains how his elongated
ellipsoid disappears:

\begin{quotation}
The length of the image of the sphere in the direction of motion is thus
greater than might otherwise be expected so that if it were not for the
flattening the sphere would appear to be elongated. \cite{6}
\end{quotation}

And also Rindler:

\begin{quotation}
This shows that a moving sphere presents a circular outline to all observers
in spite of length contraction (or rather: because of length contraction;
for without length contraction the outline would be distorted). \cite{18}
\end{quotation}

So according to Lampa-Penrose-Terrel the image of the rigid rod is, by
compensation with Einstein's contraction, a rigid rod. This enigmatic
compensation, $L_{app}=\gamma \gamma ^{-1}L_{0}=L_{0},$ might be true (the
image of a sphere, not by LT but ''by Doppler and aberration'', would be a
sphere but only for ''sufficiently small subtended solid angle'' \cite{20,
22}). However it is clear that Einstein's convention is different to
Poincar\'{e}'s one: in Poincar\'{e}'s SR the elongated \textit{light}
ellipsoid \textit{appears} because of the contraction of unit of length (see
conclusion) whilst in Einstein-Minkowski-Penrose's SR the elongated \textit{%
material }ellipsoid \textit{disappears} because of Einstein's contraction.
Let us remark that in this scientific tradition (the relativistic shape of a
sphere of matter), which begins in 1924 with Lampa, nobody made the
slightest reference to Poincar\'{e}'s 1906 elongated ellipse (the
relativistic shape of a sphere of light).

\section{Acknowledgements}

I would like to thank Jean Reignier, Thomas Durt, Pierre Marage, Nicolas
Vansteenkiste, Carmina Serrano and Marisa Serrano.

\end{document}